\documentclass[12pt]{iopart}
\begin{document}

\title[Microscopic Calculations of Nuclear EDM]
{Microscopic Calculations of Nuclear EDM}

\author{Sachiko Oshima, Takehisa Fujita and Tomoko Asaga}

\address{Department of Physics, Faculty of Science and Technology,  
Nihon University, Tokyo, Japan}

\eads{\mailto{fffujita@phys.cst.nihon-u.ac.jp} and \mailto{asaga@phys.cst.nihon-u.ac.jp}}

\begin{abstract}
We carry out the microscopic calculations of the atomic EDM 
in deuterium, Xe and Hg atomic systems. 
Due to Schiff's theorem, we obtain the atomic EDM only from 
the finite nuclear size effects, but it is found that the atomic EDM is 
appreciably larger than expected. This is essentially due to the new mechanism 
in which the nuclear excitations are taken into account while atomic states stay 
in the ground state.  The EDM of deuterium is found to be  $ d_D \simeq 0.017 d_n$ 
while the EDM of Xe and Hg become $ d_{\rm Xe} \simeq 1.6 d_n $ and 
$ d_{{\rm Hg}} \simeq -2.8 d_n $, respectively. It turns out that the new 
constraint on the neutron EDM becomes 
$ d_n \simeq (0.37\pm 0.17 \pm 0.14)\times 10^{-28} \  {\rm e}\cdot 
{\rm cm}  $. 
\end{abstract}



\maketitle

\section{Introduction}

The existence of T-violating interaction in nature is one of the most fundamental 
subjects in field theory. There must be some T-violating interaction in nature 
as long as we believe the big bang cosmology since the big bang must have started from 
the T-violating interaction. 

In order to see the violation of the T-invariance, one measures 
electric dipole moments (EDM) of particles, nuclei 
and atoms  in their ground states. Until now, the upper limit of the neutron EDM $d_n$ 
is around \cite{ne1, ne2, q1}
$$ d_n \simeq (1.9\pm 5.4)\times 10^{-26} \  {\rm e}\cdot {\rm cm} . \eqno{(1.1a)} $$ 
$$ d_n \simeq (2.6\pm 4.0 \pm 1.6)\times 10^{-26} \  {\rm e}\cdot {\rm cm} . \eqno{(1.1b)} $$ 
There have been many experimental efforts to measure the EDM of the 
atomic systems. The good examples are found for the EDM of 
$^{129}{\rm Xe}$ \cite{w22,w2} and $^{199}{\rm Hg}$ \cite{q22,q2}. 
In this case, however, one has to be careful for extracting the 
EDM of  electrons or nucleons from the 
measurement of the atomic EDM since there is Schiff's 
theorem \cite{q3}. This theorem states that the EDM of the atom is canceled out 
due to the symmetry restoration mechanism as long as the constituents are 
interacting through the electromagnetic interactions with the nonrelativistic 
kinematics. In order to obtain the EDM of atomic system, one has to find 
the relativistic effects in the EDM operators  \cite{q13,q44}. Since electrons 
become relativistic in heavy atoms, the EDM of the atomic systems becomes 
larger than the electron EDM $d_e$. 
In fact, the EDM of Cs atom has a large 
enhancement factor, namely $d_{Cs} \simeq 91 d_e$ \cite{q44,t1,t2,t3,t4,t5,t6,t7}. 

However, it is also believed that the electron EDM might well be rather 
small compared to the neutron EDM. This is based on the EDM operator 
which is derived from the supersymmetry model calculations \cite{q11,s1,s2,s3,s4,s5}. 
The EDM interaction $ {\cal H}_{edm} $ is written for $\psi_i$ fermion field 
with the EDM coupling constant $d_i$, 
$$ {\cal H}_{edm} = - {i\over 2}d_i \bar \psi_i \sigma_{\mu \nu}
 \gamma_5 \psi_i F^{\mu \nu} \eqno{(1.2)} $$
where $ F^{\mu \nu} $ denotes the electromagnetic field strength.  
Up to now, there are many efforts to determine the strength of the EDM 
coupling constant $d_i$  from the supersymmetric model calculations. 

At the present stage, however, it is not clear how large the electron EDM $d_e$ 
and neutron EDM $d_n$ should be. But most of the estimations 
of the EDM suggest that the electron EDM must be much smaller than 
the neutron EDM \cite{q13}. 
This means that it is better if one 
can measure the neutron EDM from atomic systems.

Due to the presence of Schiff's theorem, it has been long believed 
that the EDM of nuclear systems should be quite small, and 
one obtains the nuclear EDM from Schiff moments \cite{q3}. 

However, the definition of the Schiff moment does not seem to be well established. 
Sushkov, Flambaum and Khriplovich \cite{sush} first introduced the Schiff moment 
which is generated by P- and T-violating nucleon-nucleon interactions \cite{eng}, 
and in this case, the Schiff moment is not directly proportional to the individual 
EDM of nucleons. However, according to the definition of the Schiff moment 
by Khriplovich and Lamoreaux \cite{q13}, 
the Schiff moment $\bf S$ is described in terms of the individual nucleon EDM 
$d_{p,n} $ as $$ {\bf S} \sim d_{p,n} R_0^2 {\bf I} $$
where $R_0$ and ${\bf I}$ denote the radius and the total spin of the nucleus, 
respectively. 

This Schiff moment $\bf S$ can be obtained from the unitary transformation 
of the EDM Hamiltonian in terms of $\exp ({i\sum_{i=1}^A {\bf d}_N\cdot{\bf P}_i })$ 
where ${\bf P}_i$ denotes the nucleon momentum operator in nucleus.   
But we should note that this type of the unitary transformation makes the calculation 
more complicated than the calculation with the original Hamiltonian. 
This is entirely due the fact that the nuclear system is not a neutral 
system, and therefore there is no Schiff shielding in the finite size 
effects of the atomic EDM.  
Therefore, even though the unitary transformation that translates 
the nucleon coordinates by the nucleon EDM ${\bf d}_N  $ cancel out the EDM terms 
of the Hamiltonian that couple with the external electric field, 
it induces new EDM terms in the Hamiltonian some of which correspond to the Schiff moment. 
In this case, however, the translation of the nucleon coordinates affects the nuclear wave 
functions since the unitary transformation should naturally change the nuclear 
wave functions as well, and in fact, one can easily see that the EDM terms of 
the Hamiltonian  that arise from this unitary transformation can be canceled out 
by the EDM terms from the nuclear wave function effects. 
Thus, the procedure of the unitary transformation should not change the EDM evaluation 
from the one that starts from the original EDM Hamiltonian. Instead, one should be careful 
for the evaluation of the nuclear EDM with the unitary transformation. 
If one solves the eigenvalues of the transformed Hamiltonian properly, then one finds 
the same result as the one from the original Hamiltonian. 

Also, we should add a comment on Schiff's theorem why the unitary transformation can give 
a proper description of the EDM. In Schiff's theorem, the unitary transformation 
cancels out the EDM terms completely from the Hamiltonian. In this case, 
the wave functions without the EDM terms can be employed since they are the eigenstates 
of the unperturbed Hamiltonian without the EDM terms.

This point will be discussed in detail in ref. \cite{ta1}, and 
here we only note that the EDM Hamiltonian of the present treatment 
is the same as the original Hamiltonian before the unitary transformation 
of the EDM operators.  
Further, we evaluate the EDM contributions corresponding to the EDM from 
the Schiff moment. But it turns out that this EDM contribution from the Schiff 
moment is very small as will be discussed in detail later.

In this paper, we present a new calculation of the atomic EDM 
starting from the microscopic interactions in atomic and nuclear Hamiltonian. 
Here, we calculate the atomic EDM by considering all possible cases of 
the second order perturbation energies which are proportinal to the individual 
particles EDM $d_i$ and the external field $E_{ext}$.  Among many terms of 
the second order perturbation energies, the Schiff shielding can be viewed as 
the cancellation between the first order perturbation energy of $d_i E_{ext}$, 
and the second order perturbation energy of the two terms ($d_i \nabla_i V_c )$ 
and ($e{\bf r}\cdot {\bf E}_{ext} $) where $V_c$ denotes the Coulomb force 
of the corresponding system.  

The evaluations of the EDM are carried out in the three steps. 
In the first step, we take the nucleus as a point charge, and thus the finite size 
effects are neglected. In this case, we make the first and the second 
order perturbation theory and confirm that the first and the second order 
perturbation energies cancel out each other, and therefore 
there is indeed no EDM observable left due to Schiff's theorem. 
As is well known, this Schiff shielding can be obtained by the unitary 
transformation of the $total$ atomic Hamiltonian for the point nucleus case. 

In the second step, we calculate the second order perturbation theory by 
taking into account the finite nuclear size effects. In this case, however, 
we consider atomic excited states as the intermediate states, but  
the nucleus is in the ground state due to the operators involved in the processes. 
From this calculation, we obtain 
the atomic EDM ($d_A$) which can be described by  
$$ d_A \sim -{Ze^2R_0^2 \over{a_0^3\Delta E_e}}d_n 
\sim  -0.7\times 10^{-9} Z^2A^{2\over 3} d_n \eqno{(1.3a)} $$ 
for the nucleus with $Z$ protons and $A$ nucleons. 
Here $R_0$, $a_0$ and $\Delta E_e $ denote the nuclear radius, Bohr radius 
and the average electronic excitation energy, respectively. 
This result of eq.(1.3) should correspond to the part of the atomic EDM 
which arises from the Schiff moment, but the microscopic 
calculation shows that it is, in any case, extremely small. 

In the third step, we calculate the second order perturbation energy of EDM  
in nucleus, and we consider nuclear excited states as the intermediate states, 
but the electrons are in the ground state due to the operators involved 
in the processes.  This is a new mechanism that generates the second order 
perturbation energy of the atomic EDM. Indeed, at a glance, one may feel that 
one should obtain a very small effect on the EDM from the nuclear excitations 
which are always of the order of 10  MeV. However, the operators are also 
written in terms of the nuclear variables and therefore they can give rise to a normal 
nuclear perturbation result. As one knows, the physical quantities 
in the perturbation theory in nucleus are usually not very small. 
In fact, we can make a rough estimation of the atomic EDM $d_A$ as
$$ d_A \sim {Ze^2\over{\Delta E_A R_0}} d_n \sim 0.03 Zd_n \eqno{(1.3b)} $$
where we made use of the following expressions for energy denominator $\Delta E_A$ 
and nuclear radius $R_0$, 
$$ \Delta E_A \sim {41\over{A^{1/3}}} \ \ {\rm MeV}, \quad R_0 \sim 1.2 A^{1/3} 
\ \  {\rm fm}. $$
From the realistic shell model calculations, 
we obtain the EDM of the deuterium, Xe and Hg atomic systems as 
$$ d_D \simeq 0.017 d_n \eqno{(1.4a)} $$
$$ d_{\rm Xe} \simeq 1.6 d_n  \eqno{(1.4b)} $$
$$ d_{{\rm Hg}} \simeq -2.8 d_n . \eqno{(1.4c)} $$
These numbers are surprisingly large compared to eq.(1.3), and 
one sees that the nuclear EDM can be observed as the outer nucleon EDM. 
Also, the reason why one obtains 
rather large values for the heavy nucleus must be due to some 
coherence of the Coulomb interactions. Roughly, the EDM of 
the heavy nucleus is proportional to $Z$ as shown in eq.(1.3b). 

From the above numbers together with the experimental observation 
of the atomic EDM $ d_{\rm Hg}$, we obtain the most severe 
constraint on the neutron EDM $d_n$ as
$$ d_n \simeq (0.37\pm 0.17 \pm 0.14)\times 10^{-28} \  
{\rm e}\cdot {\rm cm}  \eqno{(1.5)} $$ 
where we employed the observed values of the atomic EDM  
$ d_{\rm Hg}$  \cite{q22}, 
$$ d_{\rm Hg} \simeq -(1.06\pm 0.49 \pm 0.40)\times 10^{-28} \  
{\rm e}\cdot {\rm cm} . \eqno{(1.6)} $$ 
The EDM of neutron $d_n$ in eq.(1.5) is almost three orders of magnitude smaller 
than the direct measurement of eq.(1.1), and, even more, it seems to suggest 
that the  $d_n$ is already quite close to a finite number. 

We believe that the finite number of the neutron EDM is quite 
reasonable. Since the size of neutron is $\sim 10^{-13} \ {\rm cm}$ and 
the parity violation should amount to  $\sim 10^{-6} $, therefore, 
the pure T-violation is $\sim 10^{-9} $ which is close to the baryon-photon 
ratio where the fundamental symmetries must have been violated. 

This paper is organized in the following way. In the next section, 
we treat the deuterium  case, and clarify why the nuclear perturbation 
theory can give rise to the appreciably large EDM after Schiff's shielding 
effect is removed. 
In section 3, we present a general formalism of the EDM in nuclear 
systems with atomic electrons and show that the electric dipole operators  
with nuclear variables can contribute to the atomic EDM, 
and we apply the present formalism to the Xe and Hg atomic systems. 
Also, we show the reason why the atomic EDM due to Schiff moment 
is very small.  Section 4 summarizes what we clarify in this work.

\section{EDM of deuterium}

In order to measure the EDM of atomic nuclei, we first have to face 
Schiff's theorem. The theorem states that the EDM operator 
of the particle may not have any effects on the physical observables 
since the effect of the EDM operator may be absorbed into the Coulomb 
interaction due to the translational property of the interaction Hamiltonian. 
In this case, one cannot measure any effect due to the EDM operator, 
and this is called Schiff shielding. 

However, the finite size of the nucleus brings about the EDM operator 
effects on the spectrum. In order to see the finite size effects in nucleus, 
we have to carefully write the Hamiltonian 
such that the effects can be seen explicitly. 

Here, we first write the Hamiltonian for the deuterium  system 
with explicit nuclear variables. 
The unperturbed Hamiltonian $H_0$ of the deuterium  system can be written
$$ H_0 =  { {{\bf p}}^2\over{2m} } -
{e^2\over{|{\bf r}-{\bf R}_p| }}+{ {\bf P}^2\over{M} }+V_{NN}(|{\bf R}|)  \eqno{(2.1)} $$
where $m$ and $M$ denote the masses of electron and nucleon, respectively. 
${\bf p}$ and ${\bf r}$ are the momentum and coordinate variables of 
electron while ${\bf R}_p$ and ${\bf R}_n$ denote the proton and neutron 
coordinates, respectively. ${\bf R}$ and ${\bf P}$ are the relative coordinate 
and momentum of the proton and neutron and are defined as 
$$ {\bf R}={\bf R}_p-{\bf R}_n \qquad {\bf P}={1\over 2}({\bf P}_p-{\bf P}_n) . $$
We choose the origin at the center of the deuteron, and thus we set
$$ {\bf R}_G={1\over 2} ({\bf R}_p+{\bf R}_n)=0 . $$ 
Also, $V_{NN}(|{\bf R}|)$ denotes the nucleon-nucleon interaction. 

Now, the perturbative Hamiltonian due to EDM can be written
$$ H_{edm}= - { e{\bf d}_e \cdot  ({\bf r}-{\bf R}_p )
\over{|{\bf r}-{\bf R}_p|^3}} 
-{e{\bf d}_p \cdot  ({\bf r}-{\bf R}_p )
\over{|{\bf r}-{\bf R}_p|^3}}  
-{e{\bf d}_n \cdot  ({\bf r}-{\bf R}_n )
\over{|{\bf r}-{\bf R}_n|^3}}  $$
$$+{ e{\bf d}_n \cdot {\bf R}\over{{ R}^3}} 
-({\bf d}_e+{\bf d}_p+{\bf d}_n )\cdot {\bf E}_{ext} 
+e({\bf r}-{\bf R}_p)\cdot {\bf E}_{ext}  \eqno{(2.2)} $$
where ${\bf d}_e$, ${\bf d}_p$ and ${\bf d}_n$ denote the intrinsic EDM of the electron, 
proton and neutron, respectively. They are related to the spin operator as
$$  {\bf d}_i=d_i  \mbox{\boldmath $\sigma$}_i  $$
which can be obtained by the nonrelativistic reduction of eq.(1.2). 
The last term of $H_{edm}$ comes from the normal electric dipole moment 
of the whole deuterium system which couples with the external 
electric field ${\bf E}_{ext}$. 

Eq.(2.2) can be rewritten in terms of the variable $\bf R$
$$ H_{edm}= - { e{\bf d}_e \cdot  ({\bf r}-{1\over 2}{\bf R} )
\over{|{\bf r}-{1\over 2}{\bf R}|^3}}  
-{e{\bf d}_p \cdot  ({\bf r}-{1\over 2}{\bf R} )
\over{|{\bf r}-{1\over 2}{\bf R}|^3}}  
-{e{\bf d}_n \cdot  ({\bf r}+{1\over 2}{\bf R} )
\over{|{\bf r}+{1\over 2}{\bf R}|^3}}  $$
$$+{e{\bf d}_n \cdot {\bf R}
\over{{ R}^3}} -({\bf d}_e+{\bf d}_p+{\bf d}_n )\cdot {\bf E}_{ext} 
+e\left({\bf r} -{1\over 2}{\bf R}\right)\cdot {\bf E}_{ext} . \eqno{(2.3)} $$

\subsection{ Finite size of nucleus} 

The atomic orbit is quite far from the nuclear radius, and therefore we can 
expand in terms of ${R\over{r}}$. The unperturbed Hamiltonian becomes 
$$ H_0 =  { {{\bf p}}^2\over{2m} } 
-  {e^2\over{r }} +{ {\bf P}^2\over{M} }+V_{NN}(|{\bf R}|) \eqno{(2.4a)} $$
$$ H_0^{(fs)} = -{e^2{ {\bf r}\cdot \bf R}\over{2{r}^3 }} 
+{Se^2 R^2\over{20{r}^3 }}+\cdots . \eqno{(2.4b)} $$
where $S$ in eq.(2.4b) is defined as 
$$ S={5\over 2}-{15\over 2}\cos^2 \Theta  \eqno{(2.4c)} $$
where $\Theta$ denotes the angle between the electron coordinate ${\bf r}$ 
and the nucleon coordinate ${\bf R}$. 
The perturbative Hamiltonian $H_{edm}$ can be written as the term arising 
from the point charge $H_{edm}^{(0)}$ together with the finite size Hamiltonian 
$H_{edm}^{(fs)}$ up to the order of $ (R/r)^2 $
$$ H_{edm}^{(0)} = -  {e({\bf d}_e+{\bf d}_p+{\bf d}_n) \cdot {\bf r}\over{r^3}}  
 -({\bf d}_e+{\bf d}_p+{\bf d}_n )\cdot {\bf E}_{ext} 
+e{\bf r}\cdot {\bf E}_{ext} \eqno{(2.5a)} $$
$$ H_{edm}^{(fs)} = {SeR^2\over 4r^5} ({\bf d}_e+{\bf d}_p+{\bf d}_n) 
\cdot {\bf r}    + {e{\bf d}_n \cdot {\bf R}\over{{ R}^3}}
-{1\over 2}e{\bf R}\cdot {\bf E}_{ext}  \eqno{(2.5b)} $$

\subsection{ Schiff's shielding} 

Now, we can calculate the EDM of the deuterium system from all possible cases 
of the perturbation energies up to the first order in $d_i$ and $E_{ext}$. 
We write the ground state wave function of the deuterium system as 
$$ \Psi_{De}^{(0)} = \psi_D^{(0)} ({\bf R}) \otimes \phi_e^{(0)} ({\bf r} ) \eqno{(2.6)} $$
where $\psi_D ({\bf R})$ and $\phi_e ({\bf r} ) $ denote the wave function of 
deuteron and electron, respectively. 

First, we evaluate the first order perturbation theory and obtain the EDM energy 
of the deuterium $\Delta E^{(1)}$
$$ \Delta E^{(1)} = -(d_e+d_p+d_n)E_{ext} . \eqno{(2.7)} $$
The second order perturbation theory without the finite size effect can be written as 
$$ \Delta E^{(2)}_{PC} = 
 -\sum_{n_D,n_e} {  \langle \psi_{D}^{(0)}\phi_{e}^{(0)}|  ({\bf d}_e+{\bf d}_p+{\bf d}_n)  
\cdot \nabla A_0 ({\bf r}) 
|\psi_{D}^{(n_D)}\phi_{e}^{(n_e)} \rangle \langle \psi_{D}^{(n_D)}\phi_{e}^{(n_e)}
|ez { E}_{ext}|\psi_{D}^{(0)}\phi_{e}^{(0)} \rangle \over{E_{n_D,n_e}-E_{0} }}  $$
$$ \hspace{17mm} -\sum_{n_D,n_e} {\langle\psi_{D}^{(0)}\phi_{e}^{(0)}|ez { E}_{ext}   
|\psi_{D}^{(n_D)}\phi_{e}^{(n_e)} \rangle \langle \psi_{D}^{(n_D)}\phi_{e}^{(n_e)}|
({\bf d}_e+{\bf d}_p+{\bf d}_n)  
\cdot \nabla A_0 ({\bf r})|\psi_{D}^{(0)}\phi_{e}^{(0)}\rangle\over{E_{n_D,n_e}-E_{0} }} 
\eqno{(2.8)} $$
where $\displaystyle{  A_0 ({\bf r}) ={e\over r} }$ is introduced 
and $E_{0}$ denotes the energy 
eigenvalue of the ground state in the deuterium system. 
Here, we should note that the nuclear 
part is always in the ground state and has no effect. 

Now, we make use of the following identity
$$ \nabla A_0 ({\bf r}) =i{\bf p} A_0 ({\bf r}) =
i[{\bf p},A_0 ({\bf r})]=-{i\over{e}} [{\bf p},H_0] \eqno{(2.9a)} $$
$$ H_0 |\Psi_{De}^{(0)} \rangle=E_{0} |\Psi_{De}^{(0)} \rangle, \qquad 
H_0 |\Psi_{De}^{(n)} \rangle=E_{n_D,n_e} |\Psi_{De}^{(n)} \rangle . \eqno{(2.9b)} $$
Therefore, we can sum up all the intermediate states since the energy 
denominator $E_{n_D,n_e}-E_{0} $ cancels out, and thus obtain 
$$ \Delta E^{(2)}_{PC}  = (d_e+d_p+d_n)E_{ext}   . \eqno{(2.10)} $$
This is exactly the same as the first order perturbation result with opposite 
sign, and thus it cancels out the first order perturbation result of the EDM 
in deuterium. This is just Schiff's shielding. 

\subsection{ Finite size effect (atomic excitation)} 

Now, we consider the second order perturbation theory with the finite size effects 
of the nucleus taken into account. Here, we first calculate the atomic excitation 
in the same way as the previous subsection, but consider 
the first term in eq.(2.5b) together with the last term of eq.(2.5a). 
The second order EDM energy with the finite size effect becomes
$$ \Delta E^{(2)}_{fs} =-2\sum_{n_A,n_e} {1\over{E_{n_A,n_e}-E_{0} }} $$
$$ \times  \langle \psi_D^{(0)} \phi_e^{(0)} |  {SeR^2\over4 r^5} 
({\bf d}_e+{\bf d}_p+{\bf d}_n)  \cdot  {\bf r} 
|\psi_D^{(n_D)}  \phi_e^{(n_e)} \rangle \langle\psi_D^{(n_D)}  \phi_e^{(n_e)} 
|ez { E}_{ext}| \psi_D^{(0)} \phi_e^{(0)} 
\rangle  \eqno{(2.11)}$$
where the intermediate states $|\psi_D^{(n_D)}  \phi_e^{(n_e)}  \rangle $ denote the nuclear and 
atomic excited states. From the operator properties, it is clear that 
the nuclear state should be in the ground state in the intermediate state. 

We can evaluate eq.(2.11) by making use of the closure approximation, and 
here we take $< E_{0,n_e}-E_0> \sim 10$ eV as an optimistic value. In this case, we obtain
$$ \Delta E^{(2)}_{fs} = -{e^2 \over{ 2 < E_{0,n_e}-E_0>}} (d_e+d_p+d_n) 
E_{ext} $$
$$ \times \langle\psi_D^{(0)} |SR^2|\psi_D^{(0)} \rangle\langle\phi_e^{(0)} |
{\cos^2 \theta\over{r^3}}|\phi_e^{(0)} \rangle . 
 \eqno{(2.12)}$$
We can calculate eq.(2.12), and  obtain the second order EDM 
energy and the EDM of the nucleus
$$ \Delta E^{(2)}_{fs} \simeq -1.6<S> \times 10^{-9} (d_e+d_p+d_n)E_{ext} 
 \eqno{(2.13a)} $$
$$ d_D \simeq -1.6<S> \times 10^{-9} (d_e+d_p+d_n) 
 \eqno{(2.13b)} $$
where $<S>$ denotes the expectation value of the $S$, and 
if we  calculate $<S>$ with the density distributions with the spherical
symmetry, then $<S>=0$. Therefore, the term which corresponds 
to the Schiff moment vanishes. 
Since $<S>$ is zero or of the order of unity, 
this $d_D$ is extremely small, and there is little chance to observe it. 

\subsection{ Finite size effect (nuclear excitation)} 

From eqs.(2.5), one sees that the second order perturbation energy of EDM 
has a contribution from the nuclear excitations. In this case, the atomic 
state is kept in the ground state, and we obtain the second order 
EDM energy as$$ \Delta E^{(2)}_{fs} =-2\sum_{n_D,n_e }{\langle\psi_D^{(0)} \phi_e^{(0)}  |  
{e{\bf d}_n \cdot {\bf R}\over{{ R}^3}}|\psi_D^{(n_D)} \phi_e^{(n_e)}  \rangle 
\langle  \psi_D^{(n_D)} \phi_e^{(n_e)} |(-{e\over 2} )Z{ E}_{ext}|\psi_D^{(0)} \phi_e^{(0)} 
\rangle\over{E_{n_D,n_e} -E_{0} }} \eqno{(2.14)} $$
where the intermediate electron state must be in the ground state 
due to the orthgonality condition between the ground state and the excited state since 
the operators do not depend on the electron coordinates.  

The excitations of deuteron are from continuum states. 
In this case, we can describe the state $|n_A\rangle$ and the excitation energy $E_{n_A}$ 
as
$$ |\psi_D^{(n_D)}\rangle= \frac{1}{\sqrt{V}}\exp( i{\bf k}\cdot {\bf R}) \eqno{(2.15a)} $$ 
$$ E_{n_D}={k^2\over M} . \eqno{(2.15b)} $$
Therefore, we can write eq.(2.14) as
$$ \Delta E^{(2)}_{fs} =e^2 d_nE_{ext}   
\int {d^3k\over{(2\pi)^3}}d^3Rd^3R' 
 {M\over{(k^2+k_B^2)}} \psi_D(R)\psi_D(R') 
{R\over{{R'}^2}} $$
$$ \times \cos \theta \cos \theta' \exp \left[ i{\bf k}\cdot({\bf R}-{\bf R'}) \right] 
 \eqno{(2.16)}$$
where the spin part of the deuteron wave function is already carried out. 
Also, we took only the S-state of the deuteron wave function. 
$k_B$ is related to the binding energy of the deuteron, and can be written
$$ k_B = \sqrt{ME_B} \simeq 46 \ \ {\rm MeV/c} . \eqno{(2.17)} $$
The integration over the momentum $k$ can be easily done and we obtain
$$ \Delta E^{(2)}_{fs} =e^2 d_nE_{ext} {M\over{4\pi}}  
\int d^3Rd^3R' {\exp \left[ -k_B |{\bf R}-{\bf R'}| \right] \over{|{\bf R}-{\bf R'}| }} 
 \psi_D(R)\psi_D(R') 
{R\over{{R'}^2}} 
 \cos \theta \cos \theta'  .  \eqno{(2.18)}$$
The angular integrations can be also done, and we obtain 
$$ \Delta E^{(2)}_{fs} = {4\pi\over{3}}e^2 d_nE_{ext} M k_B  
 \int R^3dRdR' \psi_D(R)\psi_D(R') $$
$$ \times\left( {\cosh (k_BR_{<})\over{k_BR_{<}}} 
-{\sinh (k_BR_{<})\over{(k_BR_{<})^2}}  \right) 
 \left( {1\over{k_BR_{>}}} +{1\over{(k_BR_{>})^2}}  \right) \exp (-k_BR_{>} )
 \eqno{(2.19)}$$
where $R_{>}$ ($R_{<}$) denotes the larger (smaller) one of $R,R'$. 

For the numerical evaluation, we use a simple deuteron wave function 
of  the following shape 
$$ \psi_D (R) = \left( {\beta^3\over{\pi}} \right)^{1\over 2} \exp (-\beta R) 
\eqno{(2.20)} $$
where we took the value of $\beta$ as $\beta = k_B =46 \ \  {\rm MeV/c} $. 
Finally, we obtain the second order EDM energy and the EDM of the nucleus
$$ \Delta E^{(2)}_{fs} \simeq 0.017 d_nE_{ext} \eqno{(2.21a)} $$
$$ d_D \simeq 0.017 d_n . \eqno{(2.21b)} $$
This is not at all a small number, and we believe that the EDM measurement 
of the deuterium system can be well competing with the neutron EDM. 

In fact, there is a proposal of EDM collaboration to measure the EDM of deuterium \cite{s6}. 
According to their proposal, they can measure the deuterium EDM up to 
$$ d_D \sim 10^{-27} \ \ {\rm e}\cdot {\rm cm} . $$
In this case, the neutron EDM $d_n$ becomes 
$$ d_n \sim 5\times 10^{-26} \ \ {\rm e}\cdot {\rm cm} $$
which is just competing with the present upper limit 
of the neutron measurement of eq.(1.1).

At this point, we should discuss the relation between the EDM [eq.(2.14)] 
due to the nuclear excitation and the EDM arising from the T-, P-odd 
nucleon-nucleon interactions. 
The T-, P-odd nucleon-nucleon interactions should be obtained from 
the second order Feynman diagram of 
the color EDM of the quark-quark-gluon vertex operator together with 
the normal quark gluon vertex terms. Therefore, it is not clear how 
one can relate this T-, P-odd nucleon-nucleon interaction to the neutron EDM.  
Further, this T-, P-odd nucleon-nucleon interactions should be 
quite complicated since they are two-body nuclear interactions. 
In this case, the calculation of the EDM should be much more involved 
than the present method, and the EDM arising 
from the T-, P-odd nucleon-nucleon interaction is 
an indirect evaluation of the atomic EDM compared to the present calculation. 
The EDM due to T-, P-odd nuclear forces is discussed by 
Flambaum, Khriplovich and Sushkov \cite{q14}, and 
they obtained the EDM of the deuteron as
$$ d_D =0.2\times 10^{-21} \xi \ \ {\rm e}\cdot {\rm cm}  $$
where $\xi$ is a parameter related to the T-, P-odd nuclear forces. 
We believe that the two different methods 
should give the different contributions to the deuteron EDM.

\section{\bf EDM of Xe and Hg atomic systems }

Up to now, we have discussed the deuterium which is the simplest 
nucleus with one electron. Now, we discuss more complicated nucleus, 
and here we treat the Xe and Hg atomic systems since there are some 
measurements of the EDM in these atomic systems \cite{w22,w2,q22,q2} and also 
there is a proposal to measure the EDM of the atomic and nuclear system \cite{q12}. 

In the same way as the deuterium case, we first write the Hamiltonian 
of the total atomic and nuclear system. 

The unperturbed Hamiltonian $H_0$ of the Xe  system can be written
$$ H_0 =  \sum_{i=1}^Z \left[ { {{\bf p}_i}^2\over{2m} } -
\sum_{j=1}^Z{e^2\over{|{\bf r}_i-{\bf R}_j| }} \right] +
{1\over 2}\sum_{i\not= j}^Z{e^2\over{|{\bf r}_i-{\bf r}_j| }} $$ 
$$ +\sum_{i=1}^A{ {\bf P}_i^2\over{2M} }+{1\over 2}\sum_{i\not= j}^A 
V_{NN}(|{\bf R}_i-{\bf R}_j|)+
{1\over 2}\sum_{i \not= j}^Z{e^2\over{|{\bf R}_i-{\bf R}_j| }}  \eqno{(3.1)} $$
where ${\bf r}_i$, ${\bf p}_i$ denote the coordinate and the momentum 
of the electron while   ${\bf R}_i$, ${\bf P}_i$ denote the nuclear variable 
and momentum, respectively. 

On the other hand, the perturbed Hamiltonian coming from the EDM is written as 
$$ H_{edm}= -  \sum_{i=1}^Z\sum_{j=1}^Z {e{\bf d}_e^i \cdot ({\bf r}_i-{\bf R}_j )
\over{|{\bf r}_i-{\bf R}_j|^3}} 
+\sum_{i=1}^Z\sum_{j\not= i}^Z {e{\bf d}_e^i \cdot ({\bf r}_i-{\bf r}_j )
\over{|{\bf r}_i-{\bf r}_j|^3}}  $$
$$ -\sum_{i=1}^Z\sum_{j=1}^A {e{\bf d}_N^j \cdot ({\bf r}_i-{\bf R}_j )
\over{|{\bf r}_i-{\bf R}_j|^3}}  
-\sum_{i=1}^A\sum_{j\not= i}^Z {e{\bf d}_N^i \cdot ({\bf R}_i-{\bf R}_j )
\over{|{\bf R}_i-{\bf R}_j|^3}}  $$
$$-\sum_{i=1}^Z{\bf d}_e^i\cdot {\bf E}_{ext} 
-\sum_{i=1}^A{\bf d}_N^i\cdot {\bf E}_{ext}
+e\sum_{i=1}^Z ({\bf r}_i-{\bf R}_i )\cdot {\bf E}_{ext}
 \eqno{(3.2)} $$
where the summation over $Z$ in nucleus means that it should be taken 
over protons. The EDM of the nucleon can be expressed in terms of the 
nucleon isospin as 
$$ {\bf d}_N^i = {1\over 2}\left[(1+\tau_i^z)d_p \mbox{\boldmath $\sigma$}^i +(1-\tau_i^z) 
d_n\mbox{\boldmath $\sigma$}^i  \right] . $$

\subsection{ Finite size of nucleus} 

Now, we evaluate the finite size effects on the second order  EDM 
energy in heavy nucleus. 
The unperturbed Hamiltonian of the atomic and nuclear system becomes 
$$ H_0 =  \sum_{i=1}^Z \left[ { {{\bf p}_i}^2\over{2m} } 
-  {Ze^2\over{r_i }} \right] +
{1\over 2}\sum_{i\not= j}^Z{e^2\over{|{\bf r}_i-{\bf r}_j| }} $$
$$ + \sum_{i=1}^A{ {\bf P}_i^2\over{2M} }
+{1\over 2}\sum_{i\not= j}^A V_{NN}(|{\bf R}_i-{\bf R}_j|)+
{1\over 2}\sum_{i\not= j}^Z{e^2\over{|{\bf R}_i-{\bf R}_j| }} . \eqno{(3.3)} $$
Here, we ignore the finite size effect of the unperturbed Hamiltonian. 

Now, the perturbed Hamiltonian $H_{edm}^{(0)}$  from the point charge and 
the Hamiltonian $H_{edm}^{(0)}$ with the finite size can be written 
up to the order of $ (R_j/r_i)^2 $
$$ H_{edm}^{(0)}= -  \sum_{i=1}^Z \left[ eZ{\bf d}_e^i\cdot {{\bf r}_i\over{r_i^3}} 
-\left(\sum_{j=1}^A e{\bf d}_N^j\right) \cdot {{\bf r}_i\over{r_i^3}}  
-  \sum_{j\not= i}^Z {e{\bf d}_e^i \cdot ({\bf r}_i-{\bf r}_j )
\over{|{\bf r}_i-{\bf r}_j|^3}} \right] $$
$$-\left(\sum_{i=1}^Z{\bf d}^i_e+\sum_{i=1}^A{\bf d}_N^i 
\right)\cdot {\bf E}_{ext} 
+e\sum_{i=1}^Z{\bf r}_i\cdot {\bf E}_{ext}  \eqno{(3.4a)} $$
$$ H_{edm}^{(fs)} = \sum_{i=1}^Z\left[ {\bf d}_e^i\cdot {\bf r}_i\sum_{j=1}^Z S_{ji}R_j^2  
+ {\bf r}_i \cdot  \sum_{j=1}^A {\bf d}_N^j S_{ji}R_j^2  \right]{e\over{r_i^5}} $$
$$ -\sum_{i=1}^A\sum_{j\not= i}^Z e{\bf d}_N^i \cdot  {({\bf R}_i-{\bf R}_j )
\over{|{\bf R}_i-{\bf R}_j|^3}}  
-{e\over 2} \sum_{i=1}^A (1+\tau_i^z) {\bf R}_i\cdot {\bf E}_{ext}  
  \eqno{(3.4b)} $$
where $S_{ji}$ in eq.(3.4b) is defined as 
$$ S_{ji}={5\over 2}-{15\over 2}\cos^2 \Theta_{ji}  \eqno{(3.4c)} $$
where $\Theta_{ji}$ denotes the angle beween the electron coordinate ${\bf r}_i$ 
and the nucleon coordinate ${\bf R}_j$, and can be given as  
$$ \cos \Theta_{ji} =\sin \theta_j \sin \theta_i \cos (\phi_j-\phi_i)
+ \cos \theta_j \cos \theta_i 
 . $$

\subsection{ Schiff shielding} 

When we treat the nucleus as a point particle, the first order 
and the second order EDM energies cancel out each other. 
That is, we first estimate the first order perturbation energy of 
the second term $ -\left(\sum_{i=1}^Z{\bf d}^i_e+\sum_{i=1}^A{\bf d}_N^i 
\right)\cdot {\bf E}_{ext} $  in eq.(3.4a), and indeed this is finite. Then, we 
evaluate the second order perturbation energy of the first term and the last term 
in eq.(3.4a). This second order perturbation energy exactly cancels out the first 
order perturbation energy, and this is indeed the Schiff schielding. 
The calculation in detail is just the same as the deuterium case, 
and therefore we do not repeat the procedure of Schiff shielding.

\subsection{ Finite size effect (atomic excitation)} 

Now, we consider the second order perturbation energy with the finite size effects 
of the nucleus. In order to evaluate the second order EDM energy, we first need to have  
the atomic and nuclear wave functions. We write the wave function of the total atomic 
and nuclear system by 
$$ \Psi_{Ae}^{(n_A,n_e)} = \psi_A^{(n_A)} ({\bf R}_1,\cdots, {\bf R}_A ) \otimes 
\phi_e^{(n_e)} ({\bf r}_1,\cdots, {\bf r}_Z ) \equiv 
\psi_A^{(n_A)}  \phi_e^{(n_e)}  \eqno{(3.5)} $$
where $n_A,\  n_e$ denote the quantum number of the excited states, and $n_A=0, \  n_e=0$ 
mean the ground states of the systems. 
Here, we assume that the atomic state has the ground state with spin zero while 
the nuclear ground state has one outer neutron with spin ${1\over 2}$. 
This is mainly because we consider the $^{129}$Xe and $^{199}$Hg 
atomic systems in this paper. 

The second order EDM energy with the finite size effect becomes 
$$ \Delta E^{(2)}_{fs} =-\sum_{n_A,n_e} {2e{ E}_{ext} \over{E_{n_A,n_e}-E_0 }} 
\langle \psi_A^{(0)}\phi_e^{(0)}  |  H_{edm}^{(fs),0}
|\psi_A^{(n_A)}  \phi_e^{(n_e)}  \rangle 
\langle \psi_A^{(n_A)}  \phi_e^{(n_e)}  |\sum_{i=1}^Z z_i 
|\psi_A^{(0)} \phi_e^{(0)} \rangle  
\eqno{(3.6)} $$
where $E_0$ denotes the ground state energy of the whole system.  
From this equation, it is clear that the intermediate nuclear state is 
the ground state $n_A=0$ due to the orthogonal condition since the operators only 
involve the electron coordinates in the last term of eq.(3.5). 
Further, $H_{edm}^{(fs),0}$ is defined as 
$$H_{edm}^{(fs),0} = \sum_{i=1}^Z\left[ {\bf d}_e^i\cdot {\bf r}_i\sum_{j=1}^Z S_{ji}R_j^2  
+ {\bf r}_i \cdot  \sum_{j=1}^A {\bf d}_N^j S_{ji}R_j^2  \right]{e\over{r_i^5}} . 
\eqno{(3.7)}  $$
Eq.(3.6) can be evaluated by employing the closure approximation
$$ \Delta E^{(2)}_{fs} \simeq -{2e^2{ E}_{ext}\over{< E_{0,n_e}-E_0>}}\times  $$ 
$$  \langle 0|\sum_{i,k=1}^Z  \left( 
({\bf d}_e^i \cdot {\bf r}^i) \sum_{j=1}^ZS_{ji}R^2_j \right.  
 + \left. \sum_{j=1}^A({\bf d}_N^j \cdot {\bf r}^i)S_{ji}R^2_j \right) 
{z_k \over{r_i^5}} |0 \rangle . \eqno{(3.8)} $$
This becomes
$$ \Delta E^{(2)}_{fs} \simeq -{2e^2Z{ E}_{ext}\over{< E_{0,n_e}-E_0>}} 
{d_n<S_{ji} \cos ^2 \theta_{i} >\langle R^2\rangle\over{a_0^3}}  \eqno{(3.9a)} $$
where the first term in eq.(3.8) vanishes since the spin of the atomic 
states is assumed to be zero. 

Here, $a_0$ denotes the Bohr radius of the atomic system and can be written as
$a_0 = {1\over Zme^2}$. 
Therefore, the atomic EDM from the atomic excitations becomes 
$$ d_A \simeq -{2e^2Z\over{mZ^2e^4}}  
{d_n<S_{ji}\cos ^2 \theta_{i} >\langle R^2\rangle\over{a_0^3}} $$ 
$$ \sim -1.0\times 10^{-9}<S_{ji}\cos ^2 \theta_{i} > Z^2 A^{2\over 3}
d_n \eqno{(3.9b)} $$
where we take $< E_{0,n_e}-E_0 > \simeq m (Ze^2)^2 $ and $\langle R^2\rangle 
\simeq r_0^2 A^{2\over 3} $ with $r_0 =1.2 $ fm. 
$<S_{ji}\cos ^2 \theta_{i} >$ is zero if we evaluate it with the density distribution 
with the spherical symmetry. 
Since $<S_{ji}\cos ^2 \theta_{i} >$ is zero or of the order of unity, 
this $d_A$ is very small and there is again little chance to observe it. 

Eq.(3.9) just corresponds to the EDM of the atomic system calculated by the Schiff 
moment in \cite{q13} even though there must be some difference by a factor of 2. 

\subsection{ Finite size effect (nuclear excitation)} 

Now, we consider the second order EDM energy due to the intermediate 
nuclear excitations. The important point is that we should estimate all possible 
cases of the second order processes with the condition that the EDM $d_i$  of 
the individual particles and the external field $E_{ext}$ must be in the first order. 
This process arises from the finite nuclear size effects in the EDM Hamiltonian. 
The second order EDM energy can be written as
$$ \Delta E^{(2)}_{fs} =-\sum_{n_A,n_e} {e^2\over{E_{n_A,n_e}-E_0 }}
\langle \psi_A^{(0)}\phi_e^{(0)} |\sum_{i=1}^A\tau_i^z {\bf R}_i\cdot {\bf E}_{ext}
  |\psi_A^{(n_A)}  \phi_e^{(n_e)}  \rangle  $$
$$ \times \langle \psi_A^{(n_A)}  \phi_e^{(n_e)} |  \sum_{i\not= j}^A 
 {1\over 4}\left[(1+\tau_i^z)d_p \mbox{\boldmath $\sigma$}^i +(1-\tau_i^z) 
d_n \mbox{\boldmath $\sigma$}^i  \right]   
 \cdot{(1+\tau_j^z)\left( {\bf R}_i-{\bf R}_j \right)
\over{|{\bf R}_i-{\bf R}_j|^3}} 
|\psi_A^{(0)}\phi_e^{(0)}  \rangle \eqno{(3.10)} $$
From eq.(3.10), it is clear that the electron state should be in the ground state ($n_e=0$) 
due to the orthogonality condition 
since the the operators in the first term in eq.(3.10) have only the nuclear coordinates. 
Here, the relation  $\displaystyle{ \sum_{i=1}^A {\bf R}_i =0 }$ should be taken 
since we set the center mass coordinate to  zero.  

Here, we carry out a rough estimation of eq.(3.10) before the realistic shell model 
calculations. Eq.(3.10) may be written as
$$ \Delta E^{(2)}_{fs} \sim {1\over \Delta E} <R>E_{ext} d_n {Ze^2\over <R^2>} \eqno{(3.11a)}  $$
Therefore, the EDM $d_A$ of the total atomic system can be written as 
$$ d_A \sim {Ze^2\over R\Delta E} d_n \sim 0.03 Z d_n \eqno{(3.11b)} $$
where we made use of the relations $\Delta E \sim \hbar \omega \sim {41\over A^{1/3} } 
\ {\rm MeV} $, $R\sim 1.2 A^{1/3} \  {\rm fm} $.

\subsection{Xe atomic system}

Now, we calculate the $^{129}{\rm Xe}$ case in which we assume a simple 
single particle shell model state, namely, $3s_{1\over 2}$ for neutrons 
together with the $2d_{3\over 2} \otimes 2^+$. 
Further, the atomic states stay in the ground state, 
and therefore the electron wave functions are not written here.  
Thus, we write the nuclear wave function as 
$$ |\Psi_{Ae}\rangle =  \sqrt{1-\alpha^2}|\nu (3s_{1\over 2})
 :{1\over 2}  \rangle +
 \alpha |\nu (2d_{3\over 2})  
\otimes 2^+ :{1\over 2} \rangle  \eqno{(3.12a)} $$ 
where $\alpha$ is a parameter which should be determined from the realistic calculations.  

In this case, the intermediate states that contribute to the second 
order EDM energy (eq.(3.9)) for $\nu (3s_{1\over 2}) $ are restricted to the following 
two states
$$ |n\rangle = |\nu (3p_{1\over 2}) :{1\over 2} \rangle \ , \ \ 
  |\nu [ (3s_{1\over 2})^2 [0^+], (2p_{1\over 2})^{-1} ]  :{1\over 2} 
\rangle  \eqno{(3.12b)} $$ 
where the second term comes from the hole state contribution. 
Here, we note that only the above two states can contribute to the matrix element 
since the matrix elements of the radial component $R$ from higher states vanish. 

Further, we assume that the $2^+$ state does not contribute to 
the atomic EDM since it is a collective state and the operators 
we are considering should not give a large contribution 
to the second order perturbation energy. This is mainly due to the fact 
that the two operators have  different rank tensors which generate $1^-$ and $0^-$. 

Therefore, we take the following intermediate states 
corresponding to the second term of eq.(3.12a) 
$$ |n\rangle = |\nu (3p_{3\over 2}) \otimes 2^+ :{1\over 2} \rangle \ , \ \  
 |\nu [(2d_{3\over 2})^2 [0^+], (2p_{3\over 2})^{-1} ]
 \otimes 2^+ :{1\over 2} \rangle . \eqno{(3.12c)} $$ 
With these simplified configurations, we can calculate eq.(3.12) and find the second order 
EDM energy due to the nuclear excitations, 
$$ \Delta E^{(2)}_{fs} ={Ze^2d_n E_{ext} \over{\omega }} 
 \left[ (1-\alpha^2) \left\{ \langle \nu (3s_{1\over 2})| Z 
|\nu (3p_{1\over 2}) \rangle 
 \langle \nu (3p_{1\over 2})|  {\mbox{\boldmath $\sigma$} 
\cdot {\bf R}\over{ R^3}} |\nu (3s_{1\over 2})  \rangle  \right. \right. $$
$$ \left. 
- \langle \nu (3s_{1\over 2})| Z 
 |\nu (2p_{1\over 2}) \rangle 
 \langle \nu (2p_{1\over 2})|  {\mbox{\boldmath $\sigma$} 
\cdot {\bf R}\over{ R^3}} |\nu (3s_{1\over 2})  \rangle \right\} $$
$$ +\alpha^2 \left\{ \langle \nu (2d_{3\over 2})\otimes 2^+ :{1\over 2} 
| Z |\nu (3p_{3\over 2})\otimes 2^+ :{1\over 2}  \rangle 
 \left.  \langle \nu (3p_{3\over 2})\otimes 2^+ :{1\over 2} 
|  {\mbox{\boldmath $\sigma$} 
\cdot {\bf R}\over{ R^3}} |\nu (2d_{3\over 2}) 
\otimes 2^+ :{1\over 2}  \rangle \right. \right. $$
$$ -\left. \langle \nu (2d_{3\over 2})\otimes 2^+ :{1\over 2} 
| Z |\nu (2p_{3\over 2})\otimes 2^+ :{1\over 2}  \rangle 
 \left.  \langle \nu (2p_{3\over 2})\otimes 2^+ :{1\over 2} 
 | {\mbox{\boldmath $\sigma$} 
\cdot {\bf R}\over{ R^3}} |\nu (2d_{3\over 2}) 
\otimes 2^+ :{1\over 2}  \rangle  \right\} \right]
\eqno{(3.13)} $$
where  $\alpha$ is taken to be $\alpha^2 \simeq 0.5$. Further, in evaluating eq.(3.10), 
we made an approximation of $ |{\bf R}_i-{\bf R}_j|^3 \approx R_i^3 $ which may 
induce some error of $20\sim 30$ percents in eq.(3.13). 

Here, we take the harmonic oscillator wave functions for the single particle 
state, and $\omega$ is given as $ \omega \simeq {41\over{A^{1\over 3}}} \ \  {\rm MeV} $. 
The matrix elements can be calculated as
$$ \langle \nu (3s_{1\over 2})| Z |\nu (3p_{1\over 2}) \rangle 
={1\over 3} \sqrt{{7\over 2M\omega}} \eqno{(3.14a)} $$
$$ \langle \nu (3p_{1\over 2})|  {\mbox{\boldmath $\sigma$} \cdot {\bf R}\over{ R^3}} 
|\nu (3s_{1\over 2})  \rangle  = \frac{157}{30 \sqrt{14 \pi}} M\omega  \eqno{(3.14b)} $$
$$ \langle \nu (2d_{3\over 2})\otimes 2^+ :{1\over 2} |
 Z |\nu (3p_{3\over 2})\otimes 2^+ :{1\over 2} \rangle 
={1\over 15} \sqrt{{2\over M\omega}} \eqno{(3.14c)} $$
$$ \langle \nu (3p_{3\over 2})\otimes 2^+ :{1\over 2} 
|  {\mbox{\boldmath $\sigma$} 
\cdot {\bf R}\over{ R^3}} |\nu (2d_{3\over 2}) 
\otimes 2^+ :{1\over 2}  \rangle 
=  \frac{5}{21}\sqrt{\frac{2}{\pi}} M\omega    \eqno{(3.14d)} $$
$$ \langle \nu (3s_{1\over 2})| Z |\nu (2p_{1\over 2}) \rangle  
= - \frac{1}{3}\sqrt{\frac{2}{M\omega}}\eqno{(3.14e)} $$
$$  \langle \nu (2p_{1\over 2})|  {\mbox{\boldmath $\sigma$} 
\cdot {\bf R}\over{ R^3}} |\nu (3s_{1\over 2})  \rangle  
= \frac{17}{15\sqrt{2\pi}} M\omega \eqno{(3.14f)} $$
$$ \langle \nu (2d_{3\over 2})\otimes 2^+ :{1\over 2} 
| Z |\nu (2p_{3\over 2})\otimes 2^+ :{1\over 2}  \rangle    
= -\frac{1}{15} \sqrt{\frac{7}{2M\omega}} \eqno{(3.14g)} $$
$$ \langle \nu (2p_{3\over 2})\otimes 2^+ :{1\over 2} 
|  {\mbox{\boldmath $\sigma$} \cdot {\bf R}\over{ R^3}} |\nu (2d_{3\over 2}) 
\otimes 2^+ :{1\over 2}  \rangle  
= \frac{22}{15} \sqrt{\frac{2}{7\pi}} M\omega .  \eqno{(3.14h)} $$

Therefore, we can evaluate eq.(3.13) and obtain $d_{{\rm Xe}} $
$$ d_{{\rm Xe}} \simeq 1.6 d_n . \eqno{(3.15)} $$
This is a surprisingly large number, but we believe that the number must be 
reliable within a factor of two or so. 

For the proton EDM, we should consider the core excitation, and 
this is rather difficult to estimate since the simple-minded 
calculation cannot give a finite number. This is simply due to the operators 
that induce the intermediate states. If we start from the core which has the $0^+$ 
state, then the dipole operator $ \tau_i^z {\bf R}_i\cdot {\bf E}_{ext} $ 
can only generate the isovector $1^-$ states while the interaction part 
${\mbox{\boldmath $\sigma$} \cdot {\bf R}\over{ R^3}} $ can only generate  
$0^-$ states. Therefore, if we start from the core with $0^+$ state, 
then there is no contribution to the atomic EDM.

\subsection{Hg atomic system}

Next, we calculate the $^{199}{\rm Hg}$ case in which we also assume a simple 
single particle shell model states as  
$$ |\Psi_{Ae}\rangle =  \sqrt{1-\alpha_1^2-\alpha_2^2} \ |\nu (3p_{1\over 2})
:{1\over 2}  \rangle 
+ \alpha_1 |\nu (2f_{5\over 2})  
\otimes 2^+ :{1\over 2} \rangle  
+ \alpha_2 |\nu (3p_{3\over 2})  
\otimes 2^+ :{1\over 2} \rangle  
\eqno{(3.16a)} $$ 
where $\alpha_1$ and $\alpha_2$ are taken to be 
$\alpha_1^2 \simeq 0.25$ and $\alpha_2^2 \simeq 0.25$. 
Here, we consider the following intermediate states for the first term of eq.(3.16a), 
$$ |n\rangle = |\nu (4s_{1\over 2})  :{1\over 2} \rangle \ , \ \ 
  |\nu [(3p_{1\over 2})^2 [0^+], 
(3s_{1\over 2})^{-1} ]  :{1\over 2} \rangle  \eqno{(3.16b)} $$
where the second term comes from the hole state contribution. 

For the second and the third terms of eq.(3.16a), we take 
$$ |n\rangle = |\nu (3d_{5\over 2}) \otimes 2^+ :{1\over 2} \rangle 
\ , \ \ |\nu [(2f_{5\over 2})^2 [0^+], (2d_{5\over 2})^{-1} ]
 \otimes 2^+ :{1\over 2} \rangle 
 \eqno{(3.16c)} $$
$$ |n\rangle = |\nu (3d_{3\over 2}) \otimes 2^+ :{1\over 2} \rangle 
\ , \ \ |\nu [(3p_{3\over 2})^2 [0^+], (2d_{3\over 2})^{-1} ]
 \otimes 2^+ :{1\over 2} \rangle 
. \eqno{(3.16d)} $$
In this case, $ \Delta E^{(2)}_{fs}$ can be written as
$$ \Delta E^{(2)}_{fs} ={Ze^2d_n E_{ext} \over{\omega }} 
  \left[ (1-\alpha^2_1-\alpha^2_2)\left\{ \langle \nu (3p_{1\over 2})| Z 
|\nu (4s_{1\over 2}) \rangle 
 \langle \nu (4s_{1\over 2})|  {\mbox{\boldmath $\sigma$} 
\cdot {\bf R}\over{ R^3}} |\nu (3p_{1\over 2})  \rangle \right. \right. $$
$$ - \left. \langle \nu (3p_{1\over 2})| Z 
|\nu (3s_{1\over 2}) \rangle 
 \langle \nu (3s_{1\over 2})|  {\mbox{\boldmath $\sigma$} 
\cdot {\bf R}\over{ R^3}} |\nu (3p_{1\over 2})  \rangle \right\} $$  
$$ +\alpha^2_1 \left\{  \langle \nu (2f_{5\over 2})\otimes 2^+ :{1\over 2}
| Z |\nu (3d_{5\over 2})\otimes 2^+ :{1\over 2}  \rangle 
 \langle \nu (3d_{5\over 2})\otimes 2^+ :{1\over 2} 
|  {\mbox{\boldmath $\sigma$} 
\cdot {\bf R}\over{ R^3}} |\nu (2f_{5\over 2}) 
\otimes 2^+ :{1\over 2}  \rangle \right.  $$
$$ - \left. \langle \nu (2f_{5\over 2})\otimes 2^+ :{1\over 2}
| Z |\nu (2d_{5\over 2})\otimes 2^+ :{1\over 2}  \rangle 
 \langle \nu (2d_{5\over 2})\otimes 2^+ :{1\over 2} 
|  {\mbox{\boldmath $\sigma$} 
\cdot {\bf R}\over{ R^3}} |\nu (2f_{5\over 2}) 
\otimes 2^+ :{1\over 2}  \rangle  \right\} $$
$$   +\alpha^2_2 \left\{ \langle \nu (3p_{3\over 2})\otimes 2^+ :{1\over 2} 
| Z |\nu (3d_{3\over 2})\otimes 2^+ :{1\over 2}  \rangle 
   \langle \nu (3d_{3\over 2})\otimes 2^+ :{1\over 2} 
|  {\mbox{\boldmath $\sigma$} 
\cdot {\bf R}\over{ R^3}} |\nu (3p_{3\over 2}) 
\otimes 2^+ :{1\over 2}  \rangle 
 \right. $$
$$ - \left. \left. \langle \nu (3p_{3\over 2})\otimes 2^+ :{1\over 2} 
| Z |\nu (2d_{3\over 2})\otimes 2^+ :{1\over 2}  \rangle 
  \langle \nu (2d_{3\over 2})\otimes 2^+ :{1\over 2} 
|  {\mbox{\boldmath $\sigma$} 
\cdot {\bf R}\over{ R^3}} |\nu (3p_{3\over 2}) 
\otimes 2^+ :{1\over 2}  \rangle 
 \right\} \right] 
 . \eqno{(3.17)} $$
Here, the matrix elements become
$$ \langle \nu (3p_{1\over 2})| Z |\nu (4s_{1\over 2}) \rangle 
=-{1\over 3} \sqrt{{3\over M\omega}} \eqno{(3.18a)} $$
$$ \langle \nu (4s_{1\over 2})|  {\mbox{\boldmath $\sigma$} \cdot {\bf R}\over{ R^3}} 
|\nu (3p_{1\over 2})  \rangle  = \frac{211}{140 \sqrt{3 \pi}} 
 M\omega  \eqno{(3.18b)} $$
$$ \langle \nu (2f_{5\over 2})\otimes 2^+ :{1\over 2} 
| Z |\nu (3d_{5\over 2})\otimes 2^+ :{1\over 2}  \rangle
=- {1\over 15} \sqrt{{2\over M\omega}} \eqno{(3.18c)} $$
$$ \langle \nu (3d_{5\over 2})\otimes 2^+ :{1\over 2} 
|  {\mbox{\boldmath $\sigma$} 
\cdot {\bf R}\over{ R^3}} |\nu (2f_{5\over 2}) 
\otimes 2^+ :{1\over 2}  \rangle 
=\frac{44}{315} \sqrt{\frac{2}{\pi}} M\omega    \eqno{(3.18d)} $$
$$ \langle \nu (3p_{3\over 2})\otimes 2^+ :{1\over 2} 
| Z |\nu (3d_{3\over 2})\otimes 2^+ :{1\over 2}  \rangle 
=-{1\over 15} \sqrt{{9\over 2M\omega}} \eqno{(3.18e)} $$
$$ \langle \nu (3d_{3\over 2})\otimes 2^+ :{1\over 2} 
|  {\mbox{\boldmath $\sigma$} 
\cdot {\bf R}\over{ R^3}} |\nu (3p_{3\over 2}) 
\otimes 2^+ :{1\over 2}  \rangle
= \frac{37}{35\sqrt{2\pi}} M\omega    \eqno{(3.18f)} $$

$$  \langle \nu (3p_{1\over 2})| Z |\nu (3s_{1\over 2}) \rangle  = \frac{1}{3}\sqrt{\frac{7}{2M\omega}}
\eqno{(3.18g)} $$
$$  \langle \nu (3s_{1\over 2})|  {\mbox{\boldmath $\sigma$} 
\cdot {\bf R}\over{ R^3}} |\nu (3p_{1\over 2})  \rangle = \frac{157}{30\sqrt{14\pi}} M\omega
\eqno{(3.18h)} $$
$$
\langle \nu (2f_{5\over 2})\otimes 2^+ :{1\over 2}
| Z |\nu (2d_{5\over 2})\otimes 2^+ :{1\over 2}  \rangle  = \frac{1}{5}\sqrt{\frac{1}{2 M\omega}} 
  \eqno{(3.18i)} $$
$$ \langle \nu (2d_{5\over 2})\otimes 2^+ :{1\over 2} 
|  {\mbox{\boldmath $\sigma$} 
\cdot {\bf R}\over{ R^3}} |\nu (2f_{5\over 2}) 
\otimes 2^+ :{1\over 2}  \rangle =  \frac{8}{21}\sqrt{\frac{2}{\pi}} M\omega  
  \eqno{(3.18j)} $$
$$ \langle \nu (3p_{3\over 2})\otimes 2^+ :{1\over 2} 
| Z |\nu (2d_{3\over 2})\otimes 2^+ :{1\over 2}  \rangle   = \frac{1}{15}\sqrt{\frac{2}{M\omega}}
\eqno{(3.18k)} $$
$$ \langle \nu (2d_{3\over 2})\otimes 2^+ :{1\over 2} 
|  {\mbox{\boldmath $\sigma$} 
\cdot {\bf R}\over{ R^3}} |\nu (3p_{3\over 2}) 
\otimes 2^+ :{1\over 2}  \rangle = \frac{5}{21} \sqrt{\frac{2}{\pi}}M\omega  . 
\eqno{(3.18l)} $$

Therefore, we can evaluate eq.(3.15) and obtain $d_{{\rm Hg}} $
$$ d_{{\rm Hg}} \simeq  -2.8 d_n . \eqno{(3.19)} $$

\subsection{Neutron EDM $d_n$}

Since we have now the relation between the atomic EDM and 
the neutron EDM $d_n$, we can deduce the constraint on the $d_n$ 
from the experimental values of $d_{{\rm Xe}}$ and $d_{{\rm Hg}}$. In fact,   
they are given as \cite{w22,w2,q22,q2}
$$ d_{\rm Xe} \simeq (0.5\pm 2.4 \pm 0.1)\times 10^{-27} \  
{\rm e}\cdot {\rm cm}  \eqno{(3.20a)} $$
$$ d_{\rm Hg} \simeq -(1.06\pm 0.49 \pm 0.40)\times 10^{-28} \  
{\rm e}\cdot {\rm cm} . \eqno{(3.20b)} $$ 
By making use of eqs.(3.15) and (3.19), we obtain the constraints on the $d_n$ 
$$ d_n \simeq (0.3\pm 1.5 \pm 0.1)\times 10^{-27} \  {\rm e}\cdot {\rm cm}  
\eqno{(3.21a)} $$ 
$$ d_n \simeq (0.37\pm 0.17 \pm 0.14)\times 10^{-28} \  {\rm e}\cdot 
{\rm cm} . \eqno{(3.21b)} $$ 
This is quite a stringent constraint on the neutron EDM $d_n$  even if 
we compare it to the direct measurement  \cite{ne1, ne2, q1} of the neutron EDM $d_n$ 
in eq.(1.1). In particular, the neutron EDM $d_n$ from the Hg measurement 
seems to suggest that the $d_n$ is getting quite close to a finite number. 
It should be very interesting if one could measure the EDM of Xe nucleus 
up to the same order as the Hg case. We can expect a finite neutron EDM. 

Since the neutron EDM $d_n$ is smaller than the value given in eq.(1.1), 
this may give rise to some serious problems for the supersymmetry 
model calculations \cite{s1} which already predict the neutron EDM $d_n$ comparable 
to the value of eq.(1.1). There must be some suppression mechanism \cite{af1}. 

\section{Conclusions} 

We have carried out the microscopic calculations of the second 
order perturbation energy of the atomic EDM for deuteron, 
Xe and Hg atomic systems. The basic point of our calculation is the new 
mechanism that generates the atomic EDM where the atomic states 
stay in the ground state while the nuclear excitations arising from 
the nuclear dipole operators are taken into account. All the calculations 
are done with the shell model wave functions even though 
we have employed a simple-minded shell model state. This may induce 
some errors of a factor of two or so, but it is for sure that 
the EDM of Xenon nucleus $d_{{\rm Xe}}$ is comparable to the neutron EDM $d_n$ itself. 

The atomic EDM that arises from the atomic excitations is related 
to the Schiff moment where the atomic EDM is proportional to the 
mean square radius of nucleons, $<R^2>$. 
It turns out, however, that the EDM due to the Schiff moment 
is very small if we evaluate it microscopically. Therefore, 
the atomic EDM from the atomic excitations is practically 
impossible to measure.  

From the present work, one sees that the atomic EDM should be 
appropriate for the experimental observation of the neutron EDM. 
Even for deuterium or D$_2$ molecule which is better since one 
gets a factor of two compared to the deuterium, the EDM of deuterium 
is a few percent of the neutron EDM. We believe that this number can 
well compete with the neutron EDM measurement which is usually 
quite difficult due to the short life time of neutron.

\ack{We thank K. Asahi for helpful discussions and comments.}

\end{document}